\documentclass[sigconf]{acmart}
\renewcommand\footnotetextcopyrightpermission[1]{} 
\usepackage{booktabs} 
\usepackage{xcolor}
\usepackage{balance}

\usepackage[utf8]{inputenc}
\usepackage[TS1,T1]{fontenc}
\usepackage{booktabs} 
\usepackage{algorithm}
\usepackage{algorithmic}
\usepackage{amsmath}
\usepackage{listings}

\newcommand{\insql}[1]{\textsf{#1}}

\begin{document}
\title{Zooming in on NYC taxi data with Portal}

\author{Julia Stoyanovich}
\affiliation{
  \institution{Drexel University\\ Philadelphia, PA}
}
\email{stoyanovich@drexel.edu}

\author{Matthew Gilbride}
\affiliation{
  \institution{Drexel University\\ Philadelphia, PA}
}
\email{mtg5014@gmail.com}

\author{Vera Zaychik Moffitt}
\affiliation{
  \institution{Drexel University\\ Philadelphia, PA}
}
\email{zaychik@drexel.edu}

\begin{abstract}

In this paper we develop a methodology for analyzing transportation data at different levels
of temporal and geographic granularity, and apply our methodology to the TLC Trip Record Dataset, made publicly available by the NYC Taxi \& Limousine Commission.  This data is naturally represented by a set of trajectories, annotated with time and with additional information such as passenger count and cost.  We analyze TLC data to identify hotspots, which point to lack of convenient public transportation options, and popular routes, which motivate ride-sharing solutions or addition of a bus route.

Our methodology is based on using a system called Portal, which implements efficient representations and principled analysis methods for evolving graphs. Portal is implemented on top of Apache Spark, a popular distributed data processing system, is inter-operable with other Spark libraries like SparkSQL, and supports sophisticated kinds of analysis of evolving graphs efficiently. Portal is currently under development in the Data, Responsibly Lab at Drexel. We plan to release Portal in the open source in Fall 2017.

\end{abstract}



\maketitle

\section{Introduction}
\label{sec:intro}

Many first-time visitors to New York City are surprised by the ubiquity of yellow cabs.  Are NYC taxis a luxury and a menace to pedestrians, bicyclists and other drivers? Or are they a necessity --- an efficient and cost-effective way to supplement public transportation options in the City that Never Sleeps, but where it can take you longer to go cross-town by subway or bus than if you were to walk, and where the only practical way to get to an airport is by taxi? 

We set out to answer these questions by analyzing the TLC Trip Record Data, made publicly available by the NYC Taxi \& Limousine Commission.\footnote{\url{http://www.nyc.gov/html/tlc/html/about/trip_record_data.shtml}} We downloaded 1 year worth of yellow cab data, spanning July 2015 through June 2016. Yellow cabs pick up passengers in Manhattan or at an airport like JFK and La Guardia, and drive them to destinations in any of NYC's five boroughs. 

\subsection{Research questions}
\label{sec:questions}

In this work, we pose the following questions:

\begin{figure*}[h!]
\centering
\includegraphics[width=4.5in]{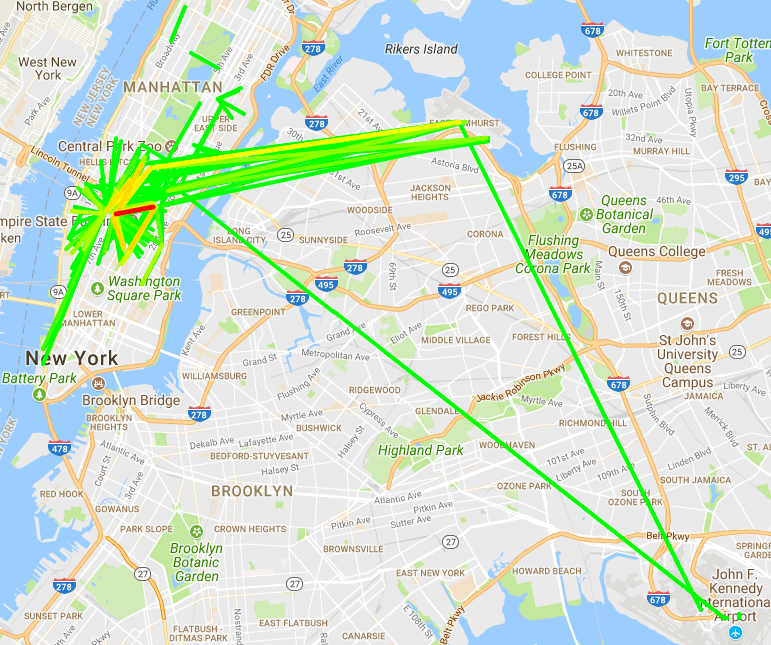}
\caption{Top-$300$ most frequent routes in March 2016, at 100-meter resolution.  Data matches that in Figure~\ref{fig:pairs_100}.}
\label{fig:map_full}
\end{figure*}

\paragraph{\bf Hotspots:}  What are the hotspots in NYC yellow cab utilization?  Which origins and destinations are the most popular? Do popularity trends persist at different levels of geographic granularity? Presence of hotspots indicates that public transportation options are insufficient in these geographic locations. 

\paragraph{\bf Popular routes:} Do there exist sets of trips that (a) share an origin and a destination, and (b)  originate at the same time, or within a few minutes of each other?  If so, a point-to-point ride-sharing solution can be implemented to reduce the cost of the trip per passenger.

\paragraph{\bf Summary of results:} We analyzed TLC data and found that there are indeed taxi utilization hotspots, and that although many popular origins and destinations persist at different levels of granularity, some do not.  For example, Penn Station, JFK airport and La Guardia (LGA) airport are among the top-$5$ locations by both out-degree (origin) and in-degree (destination) at 10-meter and 100-meter location resolution, and are clearly hotspots at 100-meters, pointing to two unsurprising facts --- that much of the taxi utilization in New York City is by tourists or by locals who travel in and out of the City, and that public transportation options serving these locations are insufficient.  Perhaps more surprising is the proportion of the total number of cab rides in NYC with JFK, La Guardia or Penn Station as their origin, destination or both.  Further, while Penn Station remains a hotspot at 1000-meter resolution, JFK and La Guardia are no longer among the top-$5$.  

We found that there exist many popular routes, some with as many as 11 simultaneous trips during a particular month in 2016, and with only 1 or 2 passengers per car --- a clear ride-sharing opportunity.  Figure~\ref{fig:map_full} gives an at-a-glance view of the results of our frequent route analysis.  This figure shows the total number of trips between pairs of locations in March 2016, for 300 most frequent such pairs.  One immediate insight from this visualization is that La Guardia participates in many more taxi trips than JFK, despite being a much smaller airport.~\footnote{\url{https://en.wikipedia.org/wiki/LaGuardia_Airport}}  Another insight is that the single most frequent route is between Penn Station and Grand Central --- two train station in Midtown Manhattan.  

Both findings can be explained by the lack of convenient public transportation options that connect these out-of-town transportation hubs:  Unlike JFK, La Guardia is not reachable by subway from Midtown Manhattan.  It takes 2 trains (with a connection at the very busy Times Square station) and 20 minutes to travel between Grand Central and Penn Station by subway.

\subsection{Context and contributions}

The research questions we ask in this paper have surely been asked and answered by others, including by ride sharing companies like Uber and Lyft, who observe passenger demand, direct cars to hotspot areas, and implement sophisticated models to adjust pricing. However, an important contribution of our work is that we present a generalizable methodology for asking these and other kinds of questions over large graphs that evolve over time.
Our methodology is based on using a system called Portal, which implements efficient representations and principled analysis methods for graphs whose topology (presence or absence of nodes and edges) and node and edge attributes change over time.  

Portal is implemented on top of Apache Spark, a popular distributed data processing system, is inter-operable with other Spark libraries like SparkSQL, and supports sophisticated kinds of analysis of evolving graphs efficiently.  
Portal implements a principled set of algebraic operations over evolving graphs~\cite{DBPL2017}, and supports concise specification of sophisticated analysis tasks: research questions of the kind we ask here can be investigated by writing a handful of lines of Portal code.  Answers to these questions can be computed in minutes for graphs with millions of edges on a modest-size cluster.

While we apply Portal to transportation data in this paper, other use cases include analysis of social and interaction networks, of biological pathways, and of knowledge bases, to name just a few.   Technically speaking,
Portal supports analytical evolutionary analysis of networks, where the goal is to model, quantify and understand the changes that have occurred in the underlying network over time~\cite{DBLP:journals/csur/AggarwalS14}.  The system supports property graphs~\cite{DBLP:journals/corr/AnglesABHRV16}, and can model and interrogate changes in network topology and in the values of node and edge attributes.  

The data analysis methodology embodied by Portal and presented here is complementary to that of TaxiViz~\cite{DBLP:journals/tvcg/FerreiraPVFS13},  a system for visual exploration of transportation data that was applied to (an earlier version of) the NYC TLC dataset.  TaxiViz supports query types in Peuquet’s spatio-temporal framework~\cite{Pequet}, interrogating data along three dimensions: space (where), time (when) and objects (what).  In contrast, our representation and query mechanisms are based on evolving graph models, rather than on spatio-temporal models. In this paper, we will showcase a particular aspect of this functionality --- temporal and structural zoom.

Portal is under development in the (\href{https://dataresponsibly.com/}{Data, Responsibly Lab})
at Drexel.  We plan to release Portal in the open source in Fall 2017.

\section{Materials and Methods}
\label{sec:methods}

\subsection{Data}
\label{sec:methods:data}

We analyzed 12 months worth of yellow cab data, spanning July 2015 through June 2016, that makes part of the TLC Trip Record dataset~\footnote{\url{http://www.nyc.gov/html/tlc/html/about/trip_record_data.shtml}}.  This dataset  lists pick-up and drop-off locations for each trip at 1 meter precision.  In our analysis, we consider locations at different levels of geographic granularity, zooming out to 10m, 100m and 1000m spatially.  In addition to pick-up and drop-off locations, the dataset also lists the pick-up and drop-off times, the number of passengers, and the fare amount. 

We represent this dataset of time-stamped trajectories with an evolving graph.  In this graph, nodes correspond to locations, and directed edges correspond to trips.  Latitude and longitude of the location is stored as an attribute of the node, while trip duration, fare and passenger count are stored as edge attributes.  We record each trip from location $a$ to location $b$ as a single directed edge from $a$ to $b$, and so our representation is technically a directed multi-graph, since multiple edges can connect a given pair of nodes.

The advantage of using an evolving graph representation for TLC data is that we can assign periods of validity to nodes and edges, and have our data manipulation operations handle these periods of validity implicitly.  We will discuss this further when explaining our analysis methods in Section~\ref{sec:methods:analysis}. We assign periods of validity to nodes and edges as follows.  An edge is valid for the duration of the trip that it represents. A node becomes valid when the first trip originates from or arrives at that node, and persists until the last incoming or outgoing trip. 

We cleaned TLC data, removing trip records that did not appear to be formatted properly, for which latitude or longitude was set to 0, or for which trip duration did not appear valid, such as when arrival time was later than departure time or when the trip took longer than 2 hours. 

\subsection{Software and execution environment}
\label{sec:methods:portal}

To analyze TLC data and answer our research questions we used Portal, a system for usable and efficient analysis of evolving graphs.  Portal is implemented in Scala as a library of Apache Spark.  We now give some technical background on Apache Spark and Portal.

Apache Spark\footnote{http://spark.apache.org/} is a distributed open-source system similar to MapReduce~\cite{Dean2004}, but based on an in-memory processing approach~\cite{Zaharia2012}.  Data in Spark is represented by Resilient Distributed Datasets (RDDs), a distributed memory abstraction for fault-tolerant computing.  All operations on RDDs are treated as a series of transformations on a collection of data partitions, such that any lost partition may be recalculated based on its lineage.  

Apache Spark provides a higher-level abstraction over MapReduce, making it easier to use for data scientists and  developers. It is open-source, and has attracted a vibrant developer community.   Spark includes a variety of data processing libraries, including SparkSQL~\cite{Armbrust2015} --- a distributed implementation of SQL, GraphX~\cite{DBLP:conf/osdi/GonzalezXDCFS14} --- a library for analysis of static graphs, as well as streaming and machine learning modules, among others. 

\begin{figure*}[h]
\centering
\includegraphics[width=4in]{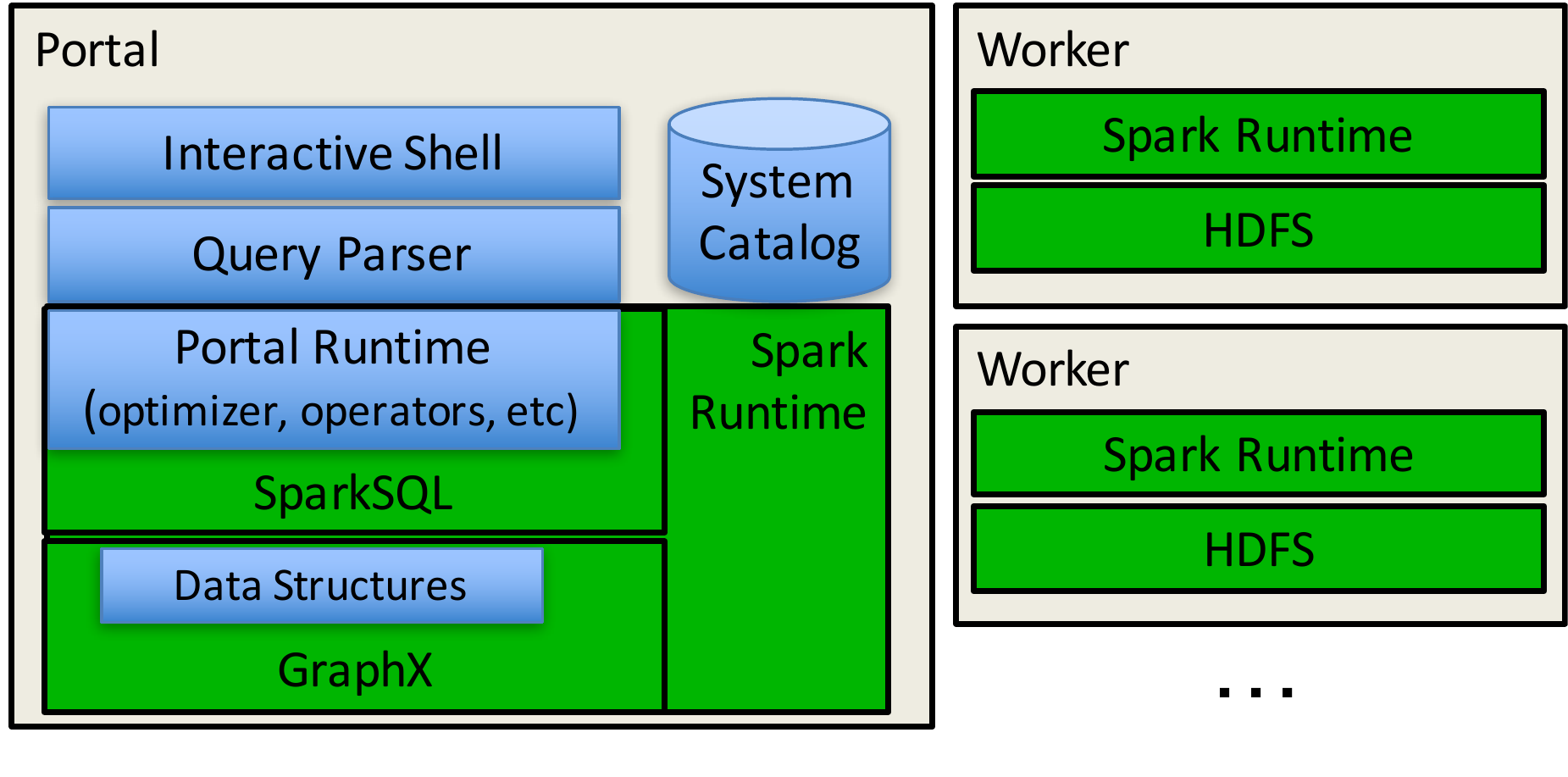}
\caption{Portal system architecture.}
\label{fig:arch}
\end{figure*}

Figure~\ref{fig:arch} shows the architecture of the Portal system on top of Apache Spark.  Green boxes indicate built-in Spark components, while blue are those we added for Portal.  Evolving graph data is distributed in partitions across a cluster of workers, is read in from Hadoop Distributed File System (HDFS)\footnote{\url{https://hadoop.apache.org/}}, and can be viewed both as an evolving graph and as a pair of RDDs.  We use the Apache Parquet format~\footnote{\url{https://parquet.apache.org/}} for on-disk representation of an evolving graphs, and provide a loader utility.

The Portal API includes the operations of TGA, an expressive temporal graph algebra we defined for querying and analysis of evolving graphs~\cite{DBPL2017}.  The API also exposes node and edge RDDs, and provides convenience methods to convert them to Spark Datasets, a relational abstraction on top of RDDs.  This feature enables inter-operability between Portal and SparkSQL.  We will showcase some TGA operations, and will demonstrate that the output of Portal can be handed over to SparkSQL for further processing,  when discussing our data analysis methodology in Section~\ref{sec:methods:analysis}. 

All data analysis was conducted on a 16-slave in-house Open Stack cloud, using Linux Ubuntu 14.04 and Spark v2.0.  Each node has 4 cores and 16 GB of RAM.  Spark Standalone cluster manager and Hadoop 2.6 were used.

\subsection{Analysis methodology}
\label{sec:methods:analysis}

We analyzed TLC data at three levels of geographic resolution: 10 meters, 100 meters (roughly 1 NYC street block) and 1000 meters, using the {\em node creation} operation.  This operation can be viewed as a generalization of the SQL \insql{group by}, applied to evolving graphs: it partitions graph nodes into non-overlapping groups based on the value of a node attribute, or on a value returned by a function that is invoked over the node.  For example, we can partition nodes that correspond to employees based on the value of the attribute \insql{department} --- all employees who work in the same department are assigned to the same group. Or we can partition nodes that represent taxi pick-up and drop-off locations by rounding up their latitude-longitude coordinates to three digits after the decimal point --- all locations that fall within the same 100m by 100m block will be assigned to the same group.

For each group of nodes in the input, a new node is created in the output. (We use Skolem functions to assign identifiers to new nodes.) The period of validity of the new node is computed by taking a temporal union of the periods of validity of the corresponding input nodes.

Edges that were present in the input persist and keep their attributes and their periods of validity as in the input, but are re-assigned to connect the newly created nodes.  For example, consider two taxi trips $e_1$ and $e_2$, both originating in the vicinity of Penn Station at locations $s_1$ and $s_2$, and both terminating at the JFK airport at locations $d_1$ and $d_2$.  If zooming out geographically places $s_1$ and $s_2$ into node group $s$, and  $d_1$ and $d_2$ into node group $d$, then edges $e_1$ and $e_2$ will both point from $s$ to $d$ in the new graph.  

We also analyzed TLC data at different levels of temporal resolution: seconds (as in the raw data), 10 minutes, 1 month and 12 months.  This was done using the {\em temporal zoom} operation, which we now describe.  Temporal zoom is analogous to node creation, in that it maps validity periods of nodes and edges of the input graph to coarser validity periods in the output.  

For example, consider two taxi trips $e_1$ and $e_2$, and suppose that they share the origin $s$ and the destination $d$, either because $s$ and $d$ were literally the same in the input dataset, or because the original locations were mapped to a pair of common locations as a result of node creation.  Suppose that $e_1$ leaves $s$ at 11:01am, and $e_2$ leaves $s$ at 11:03am.  We may want to state that $e_1$ and $e_2$ both left $s$ in the 11:01-11:10am time interval.  To do so, we invoke temporal zoom over 10-minute windows.  Under default conditions, all input nodes and edges persist in the result, but their periods of validity are adjusted (expanded) if necessary to cover the full window.  In our example above, $e_2$ will have its period of validity adjusted to start at 11:01am.

Additionally, we can specify node and edge quantifiers of the form \insql{exists} (this is the default), \insql{always} and \insql{most}, which determine under what conditions a node or an edge should be included in the temporal window. With \insql{exists}, a node or an edge is included into the window, and its validity period is adjusted, if it existed at some point during the window, even if for only one time instant.  With \insql{always}, we only include nodes and edges that existed throughout the entire window (and so an adjustment of their validity periods is unnecessary). Finally, with \insql{most}, presence during more than half of the temporal window is required.

Different levels of temporal resolution were considered for different analysis tasks, as we will discuss in detail later in this section. 

Another operation of the Portal API that was used in our analysis is {\em aggregate messages}. This operation computes the value of an attribute at a node based on information that is available at its incoming or outgoing edges, and at its immediate neighbors (nodes reachable by one edge, in either direction).  For example, we can use aggregate messages to compute the in-degree or the out-degree of a node, allowing us to quantify the number of incoming and outgoing trips for a particular location, the total cost of such trips, or their average duration.  As with other operations, time is handled implicitly.  That is, the number of incoming edges will likely differ for a given node depending on the time period (e.g., there were 35 trips out of JFK during 11:01-11:10am, and 27 trips during 11:11-11:20am), and change in this value over time is computed and handled implicitly by the system.

The Portal API supports a variety of other operations, including snapshot analytics like Page Rank, temporal and non-temporal variants of subgraph, binary operations that compute the intersection, union and difference of a pair of evolving graphs, and edge creation.  All operations implicitly handle temporal information, and also allow access to time in predicates.  Technically, TGA, the algebraic graph query language implemented by Portal, adheres to point-based temporal semantics~\cite{Bohlen2000}.  In this section, we gave a high-level overview of the operations that were used in our analysis, and refer an interested reader to~\cite{DBPL2017} for additional details.

As part of our analysis, we used node creation, temporal zoom and aggregate messages to accomplish two tasks, each corresponding to a research question in Section~\ref{sec:questions}.  We summarize these tasks here, and present results in Section~\ref{sec:results}.

{\bf Hotspots.} We analyzed the size and the degree distribution of the TLC graph: number of nodes, number of edges, and the distribution of node in-degrees (number of incoming trips) and out-degrees (number of outgoing trips).  We computed these at 10-meter, 100-meter and 1000-meter geographic resolution, with respect to the entire graph, as well as to its monthly subsets.  This was accomplished as follows:

\begin{enumerate}
\item Load the cleaned TLC dataset at 10-meter resolution.
\item Zoom out temporally to twelve 1-month windows (for per-month statistics), or to a single 12-month window (for full-graph statistics).
\item Keep original location nodes (at 10-meter resolution), or create coarser nodes at 100-meter and 1000-meter resolution.
\item Use aggregate messages to compute in-degree and out-degree of each node in each temporal window.
\item Identify hotspots: top-$k$ nodes by in-degree and out-degree.
\end{enumerate}

Hotspot analysis took between 15 and 30 minutes to execute end-to-end on a year worth of TLC data, using our in-house cluster of modest size (see Section~\ref{sec:methods:portal}).

{\bf Popular routes.} We analyzed the frequency with which multiple trips originate from the same location at the same time, and arrive at the same location. We computed these at 100-meter and 1000-meter geographic resolution, and at 10-minute temporal resolution.  At 100-meter geographic resolution, two locations are considered the same if they fall (approximately) within one NYC city block, or about 1 minute walking distance --- a reasonable distance to walk to a shared ride or to a bus.  Analysis at 1000-meter geographic resolution was done to get a sense of the general per-neighborhood trends of taxi utilization, with high utilization pointing to lack of availability of convenient public transportation options.  

Our choice of 10-minute temporal resolution means that a passenger would wait at most 10 minutes, and on average 5 minutes, for a shared ride or for a bus --- a reasonable waiting time if it leads to significant cost savings. This analysis was conducted as follows:

\begin{enumerate}
\item Load the cleaned TLC dataset at 10-meter resolution.

\item Zoom out temporally to 10-minute windows.  The effect of this operation is that all trips that start (have their pickup time) within a given 10-minute window will have their pickup time time adjusted to the beginning to the window.

\item Execute a SQL group-by query that computes, for a pair of locations $source$ and $dest$, and for a trip start time $start$, the number of originating trips, the total number of passengers, total cost of all trips, and combined trip duration.  This step is done by instantiating a Spark Dataset from the Edge RDD of the evolving graph, and passing the result to SparkSQL (see Section~\ref{sec:methods:portal}).  The following query is executed by SparkSQL at this step:

\begin{verbatim}
SELECT source, dest, start,
       count(*) as num_trips,
       sum(passengers) as total_passengers,
       sum(cost) as total_cost,
       sum(duration) as total_duration
FROM   edgeDF
GROUP BY source, dest, start
ORDER BY num_trips DESC
\end{verbatim}

\end{enumerate}

Popular routes analysis is computationally demanding because of the fine temporal granularity of the data (10 minutes).  For this analysis, we used a month worth of TLC data, corresponding to March 2016.

\section{Results}
\label{sec:results}

{\bf Hotspots.} Manhattan is dense, and the number of distinct pick-up and drop-off locations varies by about an order of magnitude as we coarsen geographic resolution by a factor of 10.  There are 1,957,882 distinct locations at 10-meter resolution, 193,676 at 100-meter resolution, and 11,540 at 1,000-meter resolution.  Our cleaned 12-month TLC dataset contains a total of 132,765,961 taxi trips between these locations.  The number of trips does not depend on the geographic resolution, and remains the same in all cases.

\begin{figure}[t!]
\centering
\begin{minipage}{3.5in}
\centering
\includegraphics[width=3.5in]{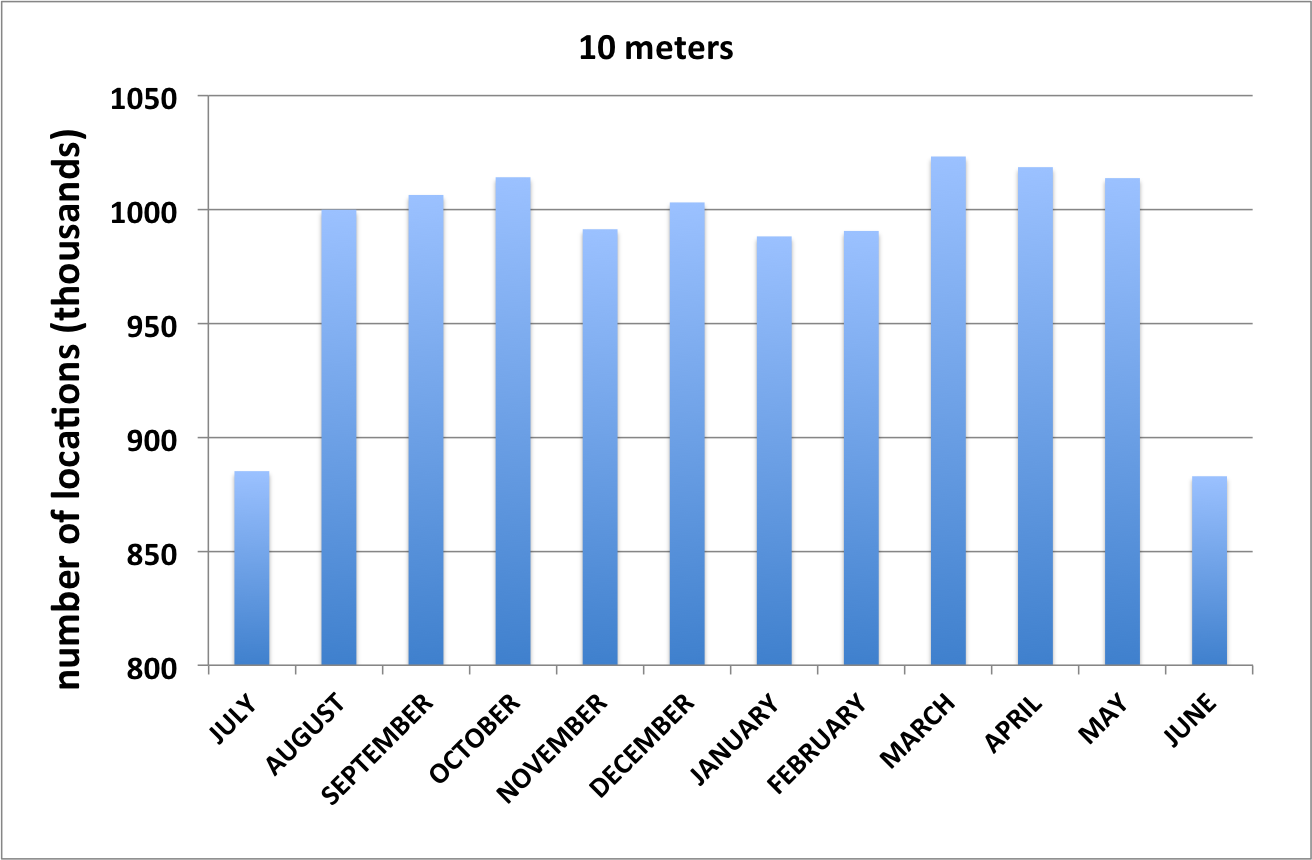}
\caption{Number of unique locations per month, at 10-meter resolution.}
\label{fig:nodes_monthly_10}\end{minipage}
\begin{minipage}{3.5in}
\centering
\includegraphics[width=3.5in]{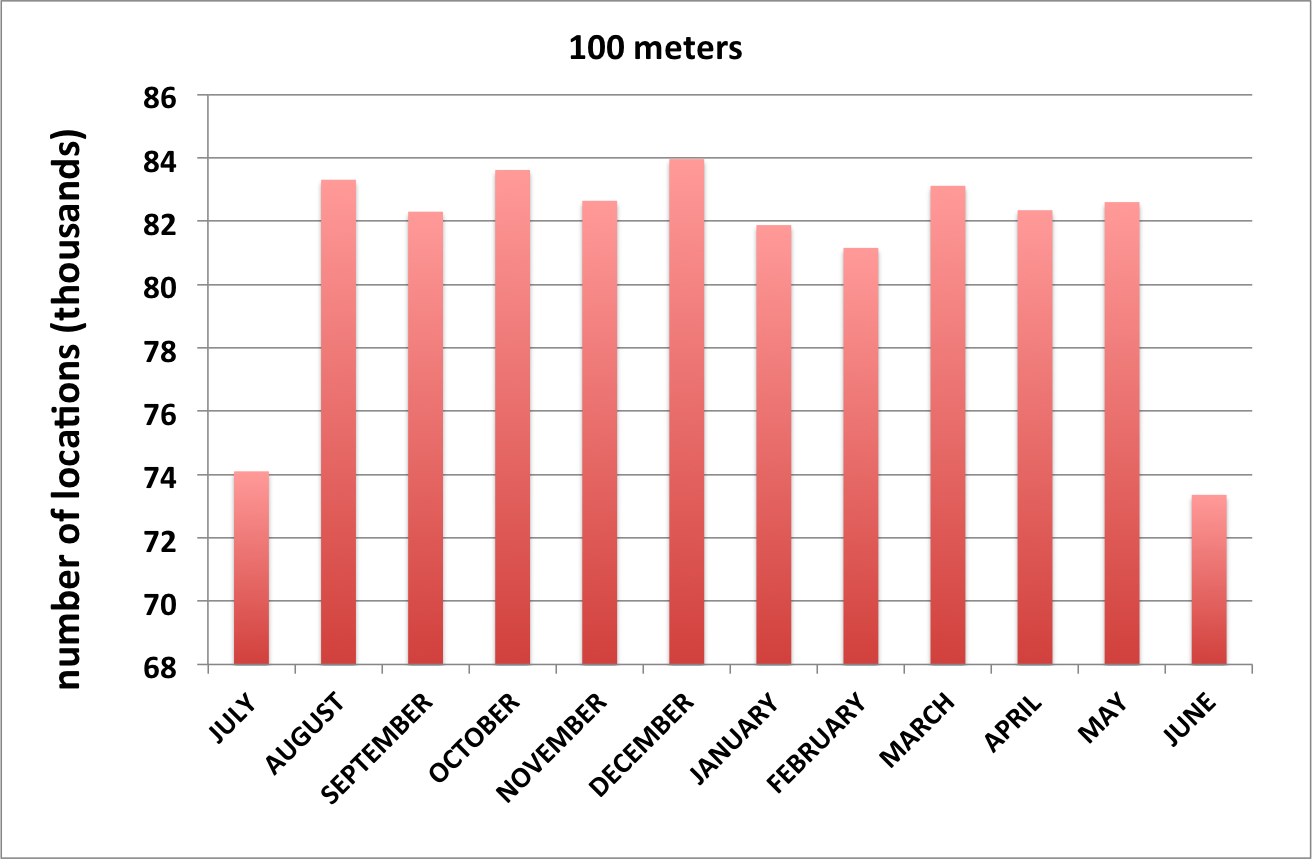}
\caption{Number of unique locations per month, at 100-meter resolution.}
\label{fig:nodes_monthly_100}
\end{minipage}
\begin{minipage}{3.5in}
\centering
\includegraphics[width=3.5in]{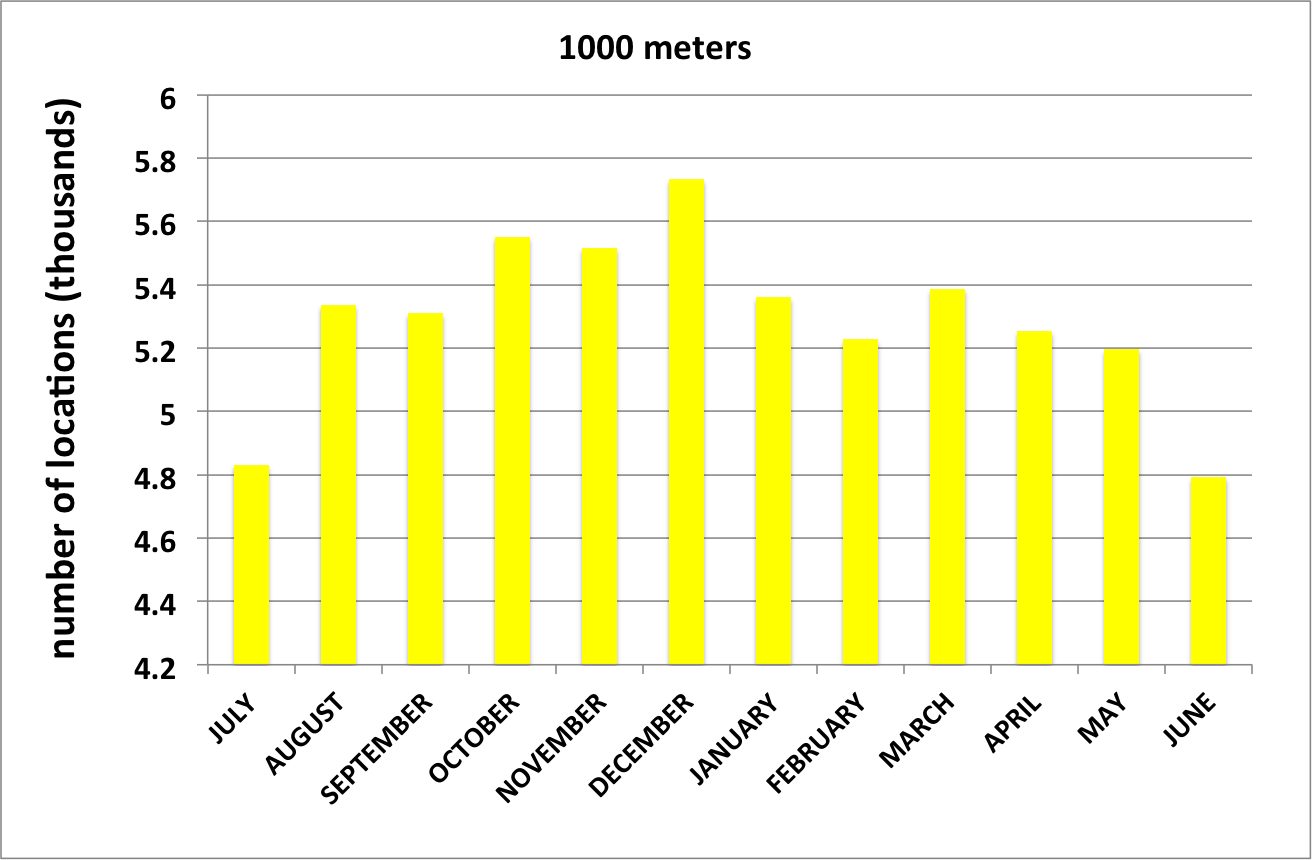}
\caption{Number of unique locations per month, at 1000-meter resolution.}
\label{fig:nodes_monthly_1000}
\end{minipage}
\end{figure}

Figures~\ref{fig:nodes_monthly_10},~\ref{fig:nodes_monthly_100} and~\ref{fig:nodes_monthly_1000} show the monthly break-down of the number of distinct locations (in thousands) at 10, 100 and 1000 meter resolutions, respectively.  We observe that the number of locations does not vary significantly month-to-month; it ranges between 882,883 and 1,023,119 for 10-meter resolution, between 74,099 and 83,115 for 100-meter resolution, and between 4793 and 5734 for 1000-meter resolution; is lowest in June and July, and highest in March for 10 and 100 meters, and in December for 1000 meters.  Month-to-month variability in the number of trips follows a similar trend as the monthly variability in the number of distinct locations.

\begin{figure}[t!]
\centering
\begin{minipage}{3.5in}
\centering
\includegraphics[width=3.5in]{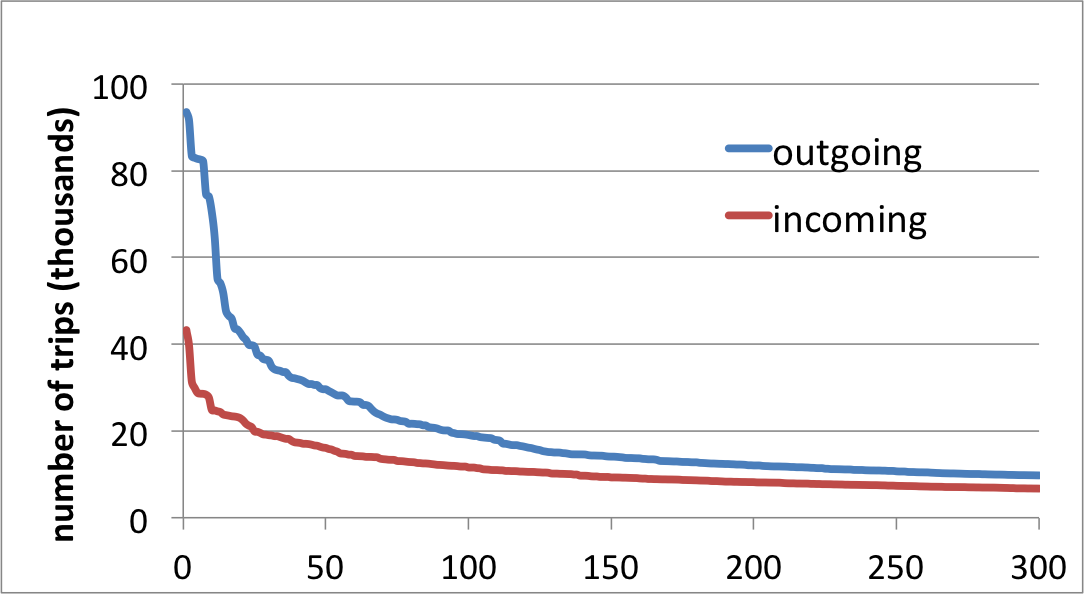}
\caption{Degree distribution for top-$300$ locations, at 10-meter resolution.}
\label{fig:degrees_10}
\end{minipage}
\begin{minipage}{3.5in}
\centering
\includegraphics[width=3.5in]{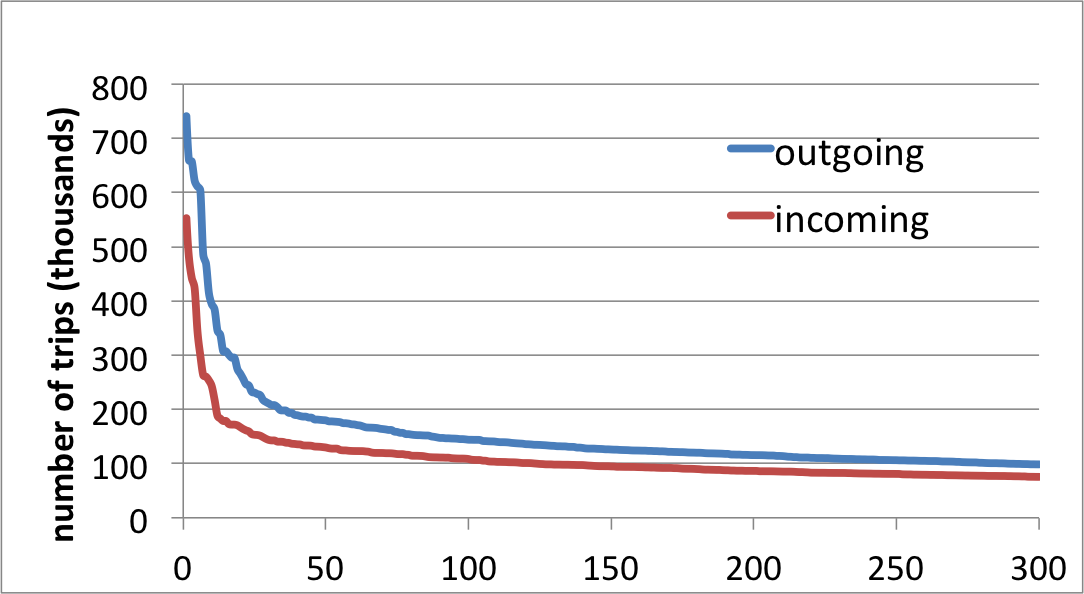}
\caption{Degree distribution for top-$300$ locations, at 100-meter resolution.}
\label{fig:degrees_100}
\end{minipage}
\begin{minipage}{3.5in}
\centering
\includegraphics[width=3.5in]{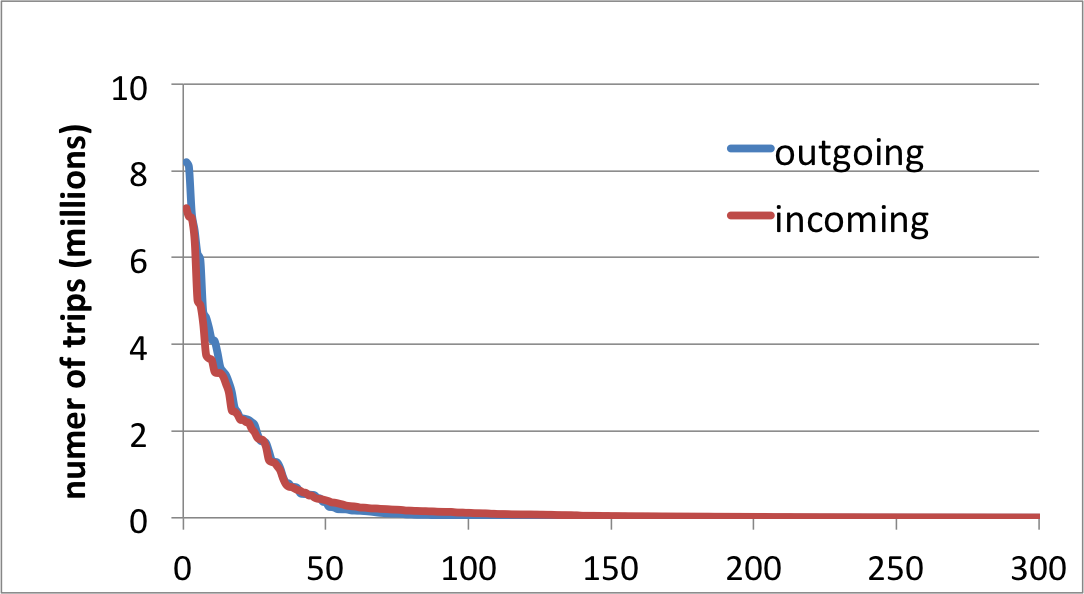}
\caption{Degree distribution for top-$300$ locations, at 1000-meter resolution.}
\label{fig:degrees_1000}
\end{minipage}
\end{figure}

\begin{figure}[t!]
\centering
\begin{minipage}{3.5in}
\centering
\includegraphics[width=3.5in]{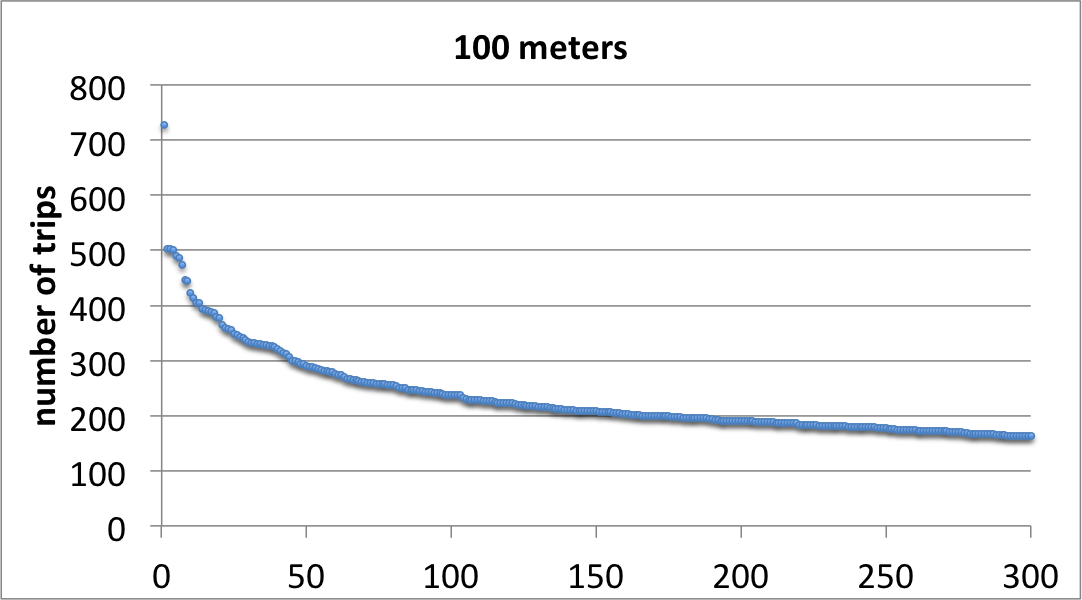}
\caption{Top-$300$ most frequent routes in March 2016, at 100-meter resolution.}
\label{fig:pairs_100}
\end{minipage}
\begin{minipage}{3.5in}
\centering
\centering
\includegraphics[width=3.5in]{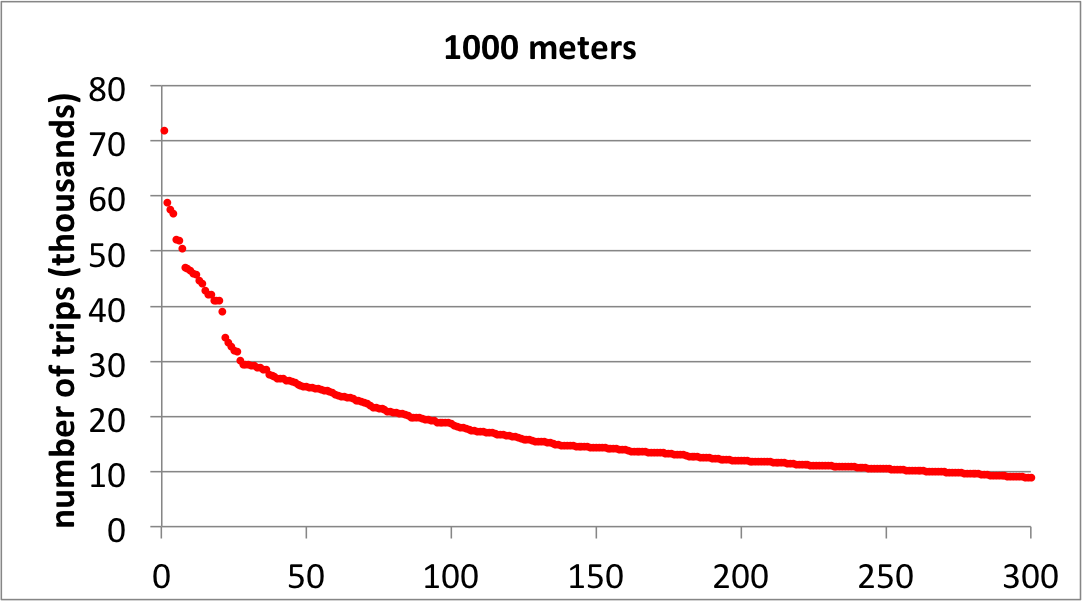}
\caption{Top-$300$ most frequent routes in March 2016, at 1000-meter resolution.}
\label{fig:pairs_1000}
\end{minipage}
\begin{minipage}{3.5in}
\centering
\includegraphics[width=3.5in]{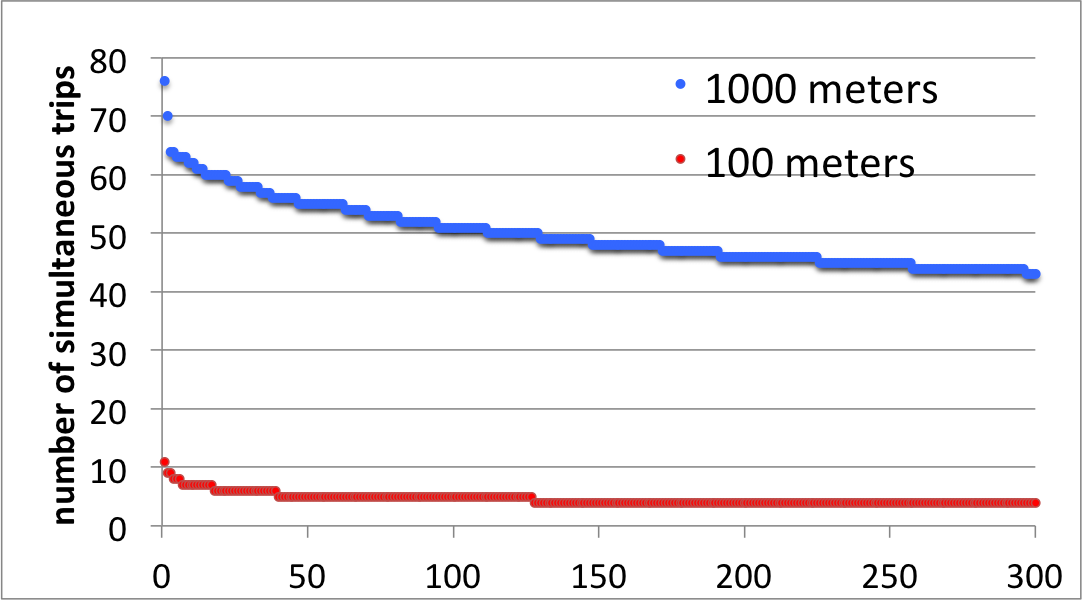}
\caption{Top-$300$ routes with highest number of simultaneous trips in March 2016, at 100-meter and 1000-meter resolution.}
\label{fig:routes_dist}
\end{minipage}
\end{figure}

\begin{figure*}
\centering
\includegraphics[width=4.5in]{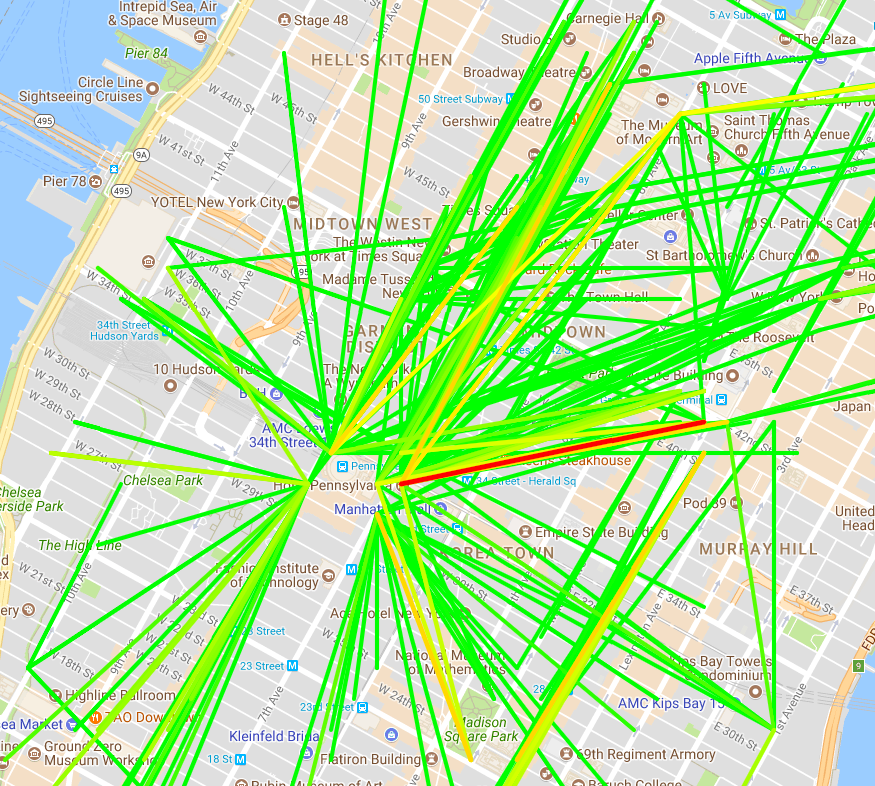}
\caption{The Midtown Manhattan portion of the top-$300$ most frequent routes in March 2016, at 100-meter resolution. Zoomed-in version of Figure~\ref{fig:map_full}, data matches that in Figure~\ref{fig:pairs_100}.}
\label{fig:map_midtown}
\end{figure*}

Figures~\ref{fig:degrees_10},~\ref{fig:degrees_100} and~\ref{fig:degrees_1000} present the distribution of node in-degrees (number of arriving trips) and out-degrees (number of originating trips) for the top-$300$ such nodes, for different geographic resolutions.  We observe that the distributions are power law, and that they are steeper for coarser resolution levels, as expected.  Even for finer geographic granularity, locations with the highest in-degree and out-degree are responsible for a very significant proportion of the edges. 

At 10-meter resolution, the top-$5$ locations by out-degree carry 434,540 edges, or about 0.3\% of all edges.  Thus, while the degree distribution at this resolution level is clearly non-uniform, hotspots are not clearly pronounced.

At 100-meter resolution, which corresponds to 1 NYC city block and is perhaps more intuitively meaningful, the top-$5$ locations have combined out-degree of 3,291,470, or 2.5\% of all edges.

At 1000-meter resolution, the top-$5$ locations by out-degree are responsible for originating 36,038,808 trips, 27\% of all taxi trips in NYC!

What are the hotspots in NYC taxi data?  Penn Station appears at the
top-5 as both an origin and a destination at all resolution levels.  In
fact, for 10-meter resolution, 3 of the 5 most frequent destinations
are in the immediate vicinity of Penn Station.  Penn Station also
appears as one of the top-5 origins at 10-meter resolution, along with
JFK (once) and La Guardia (three times) airports.  The same locations appear as hotspots at 100-meter resolution, although in a different proportion. 

Interestingly, JFK and La Guardia airports no longer appear among the
top-$5$ at 1000-meter resolution.  This can be explained by the
fact that Manhattan is dense, and while few individual city blocks
receive as much taxi traffic as an outer-borough airport, they do
jointly carry more traffic. While this finding is not altogether
surprising, it argues for the need to consider evolving graphs such as
the one we are studying in this paper at different resolution levels.

Top-$5$ origins at 1000-meter resolution include Penn Station, Midtown
East, Rockefeller Center, Bryant Park / New York Public Library, and
the Port Authority But Terminal.  Top-$5$ destinations are the same,
but Port Authority is replaced by the Flatiron district.

{\bf Popular routes.} We analyzed the frequency with which multiple trips originate from the same location at the same time, and arrive at the same location.  This analysis was done at 100-meter and 1000-meter resolution, and trips were considered to start simultaneously if they started within the same 10-minute window.  This analysis was done over one month worth of data, for the month of March 2016. 

We present summary statistics about frequent routes for March 2016.  A route is a pair of locations that serve as the origin and the destination of a taxi trip.  In our analysis in the remainder of this section, we removed routes in which origin and destination correspond to the same geographic location (by latitude / longitude, at the appropriate resolution).

Figures~\ref{fig:pairs_100} and~\ref{fig:pairs_1000} present the number of trips that took place during the month of March 2016 between a pair of locations, for 300 most frequent such pairs.  The frequencies presented in these figures consider trips that took place at any time during that month, not simultaneous trips.  

Figure~\ref{fig:map_full} presents a visualization of this data at 100-meter resolution on a map of New York City, while Figure~\ref{fig:map_midtown} zooms in on Midtown Manhattan, the area with the highest number of frequent routes.  In these visualizations, red lines correspond to the highest number of trips, while green is relatively lower.  Note, however, that even green lines connect pairs of locations with around 200 taxi trips between them (see Figure~\ref{fig:pairs_100} for a frequency break-down).

As is apparent from this visualization,  taxi trips between Penn Station and Grand Central are by far the most frequent (connected by a red line, and by multiple yellow lines).  Penn Station and Grand Central are two train stations that are in close geographic proximity: they are within a 25-minute walk from each other.  However, these hubs are not connected by a direct subway line, and so a subway trip between these two locations takes about 20 minutes and requires changing subway lines at the very busy Times Square station.

Another striking insight from the map in Figure~\ref{fig:map_midtown} is that there are many more taxi trips between Midtown Manhattan and La Guardia airport than JFK airport.  
The difference in the number of taxi trips to La Guardia vs. to JFK is particularly surprising because La Guardia is a much smaller airport: In 2016 it handled 29.8 million passengers, compared to 59.0 million at JFK.  

The most likely reason for the difference is that, unlike JFK, La Guardia is not reachable by subway.  In fact, the only public transportation option to La Guardia is a bus that runs from Upper Manhattan. While ride-sharing solutions and convenient public transportation options will not meet the needs of all travelers, having these options will help alleviate traffic congestion and reduce transportation cost for many, as suggested by the comparatively lower JFK taxi utilization.

Finally, we analyzed the number of simultaneous frequent routes: pairs of locations that were connected by multiple simultaneous taxi trips.  At 100-meter resolution, there were 55,401 routes with two or more simultaneous taxi trips. Of these, 2,049 routes had 3 simultaneous trips at some point during that month, 262 had 4 simultaneous trips, and one source-destination pair had as many as 11 simultaneous trips.  At 1000-meter resolution, the two most frequent routes had 76 and 70 simultaneous trips at some point during March 2016. Figure~\ref{fig:routes_dist} presents the number of simultaneous trips for 300 most frequent routes at 100-meter and 1000-meter resolution.  Based on our analysis, the vast majority of popular routes are taken by cabs with 1 or 2 passengers --- a clear ride-sharing opportunity!  Ride sharing is warranted particularly for the routes with 3 or more simultaneous trips.

\section{Outlook}
\label{sec:outlook}

In this paper we presented an exploration of the NYC TLC yellow cab dataset, and identified hotspots and frequent routes.  We showcased analysis methods that consider data at different levels of temporal and geographic resolution using a tool called Portal.  We plan to analyze this data further, determining persistence of hotspots and of frequent routes over time, and considering daily, weekly and seasonal trends.  We also plan to integrate taxi data with public transportation data, and to validate our methods on datasets from cities other than New York.
\balance

\bibliographystyle{ACM-Reference-Format}
\bibliography{temporal}


\begin{thebibliography}{00}


\ifx \showCODEN    \undefined \def \showCODEN     #1{\unskip}     \fi
\ifx \showDOI      \undefined \def \showDOI       #1{{\tt DOI:}\penalty0{#1}\ }
  \fi
\ifx \showISBNx    \undefined \def \showISBNx     #1{\unskip}     \fi
\ifx \showISBNxiii \undefined \def \showISBNxiii  #1{\unskip}     \fi
\ifx \showISSN     \undefined \def \showISSN      #1{\unskip}     \fi
\ifx \showLCCN     \undefined \def \showLCCN      #1{\unskip}     \fi
\ifx \shownote     \undefined \def \shownote      #1{#1}          \fi
\ifx \showarticletitle \undefined \def \showarticletitle #1{#1}   \fi
\ifx \showURL      \undefined \def \showURL       #1{#1}          \fi
\providecommand\bibfield[2]{#2}
\providecommand\bibinfo[2]{#2}
\providecommand\natexlab[1]{#1}
\providecommand\showeprint[2][]{arXiv:#2}

\bibitem[\protect\citeauthoryear{Aggarwal and Subbian}{Aggarwal and
  Subbian}{2014}]%
        {DBLP:journals/csur/AggarwalS14}
\bibfield{author}{\bibinfo{person}{Charu~C. Aggarwal} {and}
  \bibinfo{person}{Karthik Subbian}.} \bibinfo{year}{2014}\natexlab{}.
\newblock \showarticletitle{Evolutionary Network Analysis}.
\newblock \bibinfo{journal}{{\em {ACM} Comput. Surv.\/}} \bibinfo{volume}{47},
  \bibinfo{number}{1} (\bibinfo{year}{2014}), \bibinfo{pages}{10:1--10:36}.
\newblock
\showDOI{%
\url{http://dx.doi.org/10.1145/2601412}}


\bibitem[\protect\citeauthoryear{Angles, Arenas, Barcel{\'{o}}, Hogan, Reutter,
  and Vrgoc}{Angles et~al\mbox{.}}{2016}]%
        {DBLP:journals/corr/AnglesABHRV16}
\bibfield{author}{\bibinfo{person}{Renzo Angles}, \bibinfo{person}{Marcelo
  Arenas}, \bibinfo{person}{Pablo Barcel{\'{o}}}, \bibinfo{person}{Aidan
  Hogan}, \bibinfo{person}{Juan~L. Reutter}, {and} \bibinfo{person}{Domagoj
  Vrgoc}.} \bibinfo{year}{2016}\natexlab{}.
\newblock \showarticletitle{Foundations of Modern Graph Query Languages}.
\newblock \bibinfo{journal}{{\em CoRR\/}}  \bibinfo{volume}{abs/1610.06264}
  (\bibinfo{year}{2016}).
\newblock
\showURL{%
\url{http://arxiv.org/abs/1610.06264}}


\bibitem[\protect\citeauthoryear{Armbrust, Xin, Lian, Huai, Liu, Bradley, Meng,
  Kaftan, Franklin, Ghodsi, and Zaharia}{Armbrust et~al\mbox{.}}{2015}]%
        {Armbrust2015}
\bibfield{author}{\bibinfo{person}{Michael Armbrust},
  \bibinfo{person}{Reynold~S Xin}, \bibinfo{person}{Cheng Lian},
  \bibinfo{person}{Yin Huai}, \bibinfo{person}{Davies Liu},
  \bibinfo{person}{Joseph~K Bradley}, \bibinfo{person}{Xiangrui Meng},
  \bibinfo{person}{Tomer Kaftan}, \bibinfo{person}{Michael~J Franklin},
  \bibinfo{person}{Ali Ghodsi}, {and} \bibinfo{person}{Matei Zaharia}.}
  \bibinfo{year}{2015}\natexlab{}.
\newblock \showarticletitle{{Spark SQL : Relational Data Processing in Spark}}.
  In \bibinfo{booktitle}{{\em SIGMOD'15}}. \bibinfo{address}{Melbourne,
  Australia}.
\newblock
\showISBNx{9781450327589}


\bibitem[\protect\citeauthoryear{B{\"{o}}hlen, Jensen, and
  Snodgrass}{B{\"{o}}hlen et~al\mbox{.}}{2000}]%
        {Bohlen2000}
\bibfield{author}{\bibinfo{person}{Michael~H B{\"{o}}hlen},
  \bibinfo{person}{Christian~S Jensen}, {and} \bibinfo{person}{Richard~T
  Snodgrass}.} \bibinfo{year}{2000}\natexlab{}.
\newblock \showarticletitle{{Temporal Statement Modifiers}}.
\newblock \bibinfo{journal}{{\em ACM Transactions on Database Systems\/}}
  \bibinfo{volume}{25}, \bibinfo{number}{4} (\bibinfo{year}{2000}),
  \bibinfo{pages}{407--456}.
\newblock


\bibitem[\protect\citeauthoryear{Dean and Ghemawat}{Dean and Ghemawat}{2004}]%
        {Dean2004}
\bibfield{author}{\bibinfo{person}{Jeffrey Dean} {and} \bibinfo{person}{Sanjay
  Ghemawat}.} \bibinfo{year}{2004}\natexlab{}.
\newblock \showarticletitle{{MapReduce: Simplied Data Processing on Large
  Clusters}}.
\newblock \bibinfo{journal}{{\em Proceedings of 6th Symposium on Operating
  Systems Design and Implementation\/}} (\bibinfo{year}{2004}),
  \bibinfo{pages}{137--149}.
\newblock
\showISBNx{9781595936868}
\showISSN{00010782}
\showDOI{%
\url{http://dx.doi.org/10.1145/1327452.1327492}}
\showeprint[arxiv]{10.1.1.163.5292}


\bibitem[\protect\citeauthoryear{Ferreira, Poco, Vo, Freire, and
  Silva}{Ferreira et~al\mbox{.}}{2013}]%
        {DBLP:journals/tvcg/FerreiraPVFS13}
\bibfield{author}{\bibinfo{person}{Nivan Ferreira}, \bibinfo{person}{Jorge
  Poco}, \bibinfo{person}{Huy~T. Vo}, \bibinfo{person}{Juliana Freire}, {and}
  \bibinfo{person}{Cl{\'{a}}udio~T. Silva}.} \bibinfo{year}{2013}\natexlab{}.
\newblock \showarticletitle{Visual Exploration of Big Spatio-Temporal Urban
  Data: {A} Study of New York City Taxi Trips}.
\newblock \bibinfo{journal}{{\em {IEEE} Trans. Vis. Comput. Graph.\/}}
  \bibinfo{volume}{19}, \bibinfo{number}{12} (\bibinfo{year}{2013}),
  \bibinfo{pages}{2149--2158}.
\newblock
\showDOI{%
\url{http://dx.doi.org/10.1109/TVCG.2013.226}}


\bibitem[\protect\citeauthoryear{Gonzalez, Xin, Dave, Crankshaw, Franklin, and
  Stoica}{Gonzalez et~al\mbox{.}}{2014}]%
        {DBLP:conf/osdi/GonzalezXDCFS14}
\bibfield{author}{\bibinfo{person}{Joseph~E. Gonzalez},
  \bibinfo{person}{Reynold~S. Xin}, \bibinfo{person}{Ankur Dave},
  \bibinfo{person}{Daniel Crankshaw}, \bibinfo{person}{Michael~J. Franklin},
  {and} \bibinfo{person}{Ion Stoica}.} \bibinfo{year}{2014}\natexlab{}.
\newblock \showarticletitle{{GraphX}: Graph Processing in a Distributed
  Dataflow Framework}. In \bibinfo{booktitle}{{\em 11th {USENIX} Symposium on
  Operating Systems Design and Implementation, {OSDI} '14, Broomfield, CO, USA,
  October 6-8, 2014.}} \bibinfo{pages}{599--613}.
\newblock
\showURL{%
\url{https://www.usenix.org/conference/osdi14/technical-sessions/presentation/gonzalez}}


\bibitem[\protect\citeauthoryear{Moffitt and Stoyanovich}{Moffitt and
  Stoyanovich}{2017}]%
        {DBPL2017}
\bibfield{author}{\bibinfo{person}{Vera~Zaychik Moffitt} {and}
  \bibinfo{person}{Julia Stoyanovich}.} \bibinfo{year}{2017}\natexlab{}.
\newblock \showarticletitle{Evolving Graph Algebra}. In
  \bibinfo{booktitle}{{\em The 16th International Symposium on Database
  Programming Languages (DBPL)}}.
\newblock


\bibitem[\protect\citeauthoryear{Peuquet}{Peuquet}{1994}]%
        {Pequet}
\bibfield{author}{\bibinfo{person}{D. Peuquet}.}
  \bibinfo{year}{1994}\natexlab{}.
\newblock \showarticletitle{It`s about time: A conceptual framework for the
  representation of temporal dynamics in geographic information systems}.
\newblock \bibinfo{journal}{{\em {Annals of the Association of American
  Geographers}\/}} \bibinfo{volume}{84}, \bibinfo{number}{3}
  (\bibinfo{year}{1994}), \bibinfo{pages}{441–461}.
\newblock


\bibitem[\protect\citeauthoryear{Zaharia, Chowdhury, Das, Dave, Ma, Mccauley,
  Franklin, Shenker, and Stoica}{Zaharia et~al\mbox{.}}{2012}]%
        {Zaharia2012}
\bibfield{author}{\bibinfo{person}{Matei Zaharia}, \bibinfo{person}{Mosharaf
  Chowdhury}, \bibinfo{person}{Tathagata Das}, \bibinfo{person}{Ankur Dave},
  \bibinfo{person}{Justin Ma}, \bibinfo{person}{Murphy Mccauley},
  \bibinfo{person}{Michael~J Franklin}, \bibinfo{person}{Scott Shenker}, {and}
  \bibinfo{person}{Ion Stoica}.} \bibinfo{year}{2012}\natexlab{}.
\newblock \showarticletitle{{Resilient Distributed Datasets : A Fault-Tolerant
  Abstraction for In-Memory Cluster Computing}}. In \bibinfo{booktitle}{{\em
  NSDI}}.
\newblock


\end{thebibliography}

\end{document}